# Introduction of an Assistance System to Support Domain Experts in Programming Low-code to Leverage Industry 5.0

E.-M. Neumann[1], B. Vogel-Heuser[1,2], *Senior Member, IEEE*, F. Haben[1], M. Krüger[1], and T. Wieringa[3]

*Abstract*— The rapid technological leaps of Industry 4.0 increase the pressure and demands on humans working in automation, which is one of the main motivators of Industry 5.0. In particular, automation software development for mechatronic systems becomes increasingly challenging, as both domain knowledge and programming skills are required for high-quality, maintainable software. Especially for small companies from automation and robotics without dedicated software engineering departments, domain-specific low-code platforms become indispensable that enable domain experts to develop code intuitively using visual programming languages, e.g., for tasks such as retrofitting mobile machines. However, for extensive functionalities, visual programs may become overwhelming due to the scaling-up problem. In addition, the ever-shortening time-to-market increases the time pressure on programmers. Thus, an assistance system concept is introduced that can be implemented by low-code platform suppliers based on combining data mining and static code analysis. Domain experts are supported in developing low-code by targeted recommendations, metric-based complexity measurement, and reducing complexity by encapsulating functionalities. The concept is implemented for the industrial low-code platform HAWE eDesign to program hydraulic components in mobile machines, and its benefits are confirmed in a user study and an industrial expert workshop.

## I. MOTIVATION AND INTRODUCTION

To meet the growing demands on humans to adapt to the ever shorter innovation cycles of Industry 4.0, Industry 5.0 puts the human into the center of technological progress by enlarging Industry 4.0 with ecological, sustainable, and human factors [1, 2]. An increasing part of mechatronic system functionality is implemented via automation software, causing a complexity shift towards software. The shorter time-to-market increases the pressure on humans programming the software, and, thus, the importance of innovative approaches in the context of Industry 5.0 to consider human factors in software engineering is increasing. Programming automation software requires in-depth domain knowledge to consider process-specific characteristics and avoid inconsistencies. However, domain experts often do not have sound software engineering skills to implement sophisticated functionalities efficiently with high quality. Therefore, *visual programming languages (VPLs)* are widely used in computer science and mechatronics, e.g., Simulink [3] or the graphical languages of IEC 61131-3 [4] used in manufacturing systems to support the programmer (i.e., the *user*) by graphic elements, e.g., encapsulated function blocks (referred to as *blocks* as follows), arrows, or spatial separation, rather than pure text [5]. Thus, VPL support the idea of human-centered approaches in Industry 5.0. by helping users to complete tasks faster and simplify human-machine interaction and debugging [6].

With *low-code platforms*, a large application area for VPL has emerged in recent years, which abstract programs to make them understandable to technicians with extensive domain knowledge but only superficial programming skills. Low-code platforms like Siemens *Mendix* [7] or Lego *NXT-G* [8] allow rapid development of programs with as little code as possible written by hand, thus enabling faster development times and reducing software complexity [9, 10]. However, VPL and low-code platforms can have the opposite effect for extensive functionalities: Graphical elements may inflate a project, leading to the so-called *scaling up problem* [11].

To leverage Industry 5.0 and support domain experts in programming low-code, this letter presents an assistance system with three enablers: To address the scaling-up problem, the complexity of VPL software is made explicit using *software metrics* and reduced by *encapsulating recurring code artifacts*. To counter the increasing time pressure on programmers, *recommendations* for blocks to use are provided. Using the industrial low-code platform HAWE *eDesign* [12], the benefits of the assistance are demonstrated for the use case of programming hydraulic components – a highly relevant application in automation and robotics [13], where hydraulics are in widespread use to quickly generate enormous forces, e.g., in mobile machines, hydraulic presses,

Manuscript received: February 28, 2022; Revised: June 17, 2022; Accepted: July 13, 2022

This paper was recommended for publication by Ashis Banerjee upon evaluation of the Associate Editor and Reviewers' comments.

This work was supported by the Bavarian Ministry of Economic Affairs, Energy and Technology via the project AIValve (Grant No. DIK0116/01).

E. M. Neumann, B. Vogel-Heuser, F. Haben and M. Krüger are with the Institute of Automation and Information Systems, Department of Mechanical Engineering, School of Engineering and Design, Technical University of Munich, Germany, {eva-maria.neumann; vogel-heuser; fabian.haben; marius.krueger}@tum.de.

B. Vogel-Heuser is Core Member of MDSI and Member of MIRMI

T. Wieringa is with HAWE Hydraulik SE, Germany, t.wieringa@hawe.de.

Digital Object Identifier (DOI): see top of this page.

or exoskeletons. Users of such platforms are often small companies without dedicated software departments and, thus, without in-depth programming competencies to efficiently comprehend and adapt software in more complex VPL such as graphical IEC 61131-3 languages in short time. Nevertheless, domain experts have to perform an increasing amount of programming tasks in their daily work, e.g., to integrate hydraulic components into existing machines, making low-code platforms indispensable. Platform users often only have a small amount of highly application-specific software projects. Assuming that instead, platform *suppliers* usually have access to a high number of user projects, the assistance system is implemented with the supplier HAWE providing more than 1200 anonymized user projects as a data basis.

The requirements for the assistance are derived in Sec. II. Sec. III introduces the state of the art. The assistance system concept follows in Sec. IV and the implementation in Sec. V to be evaluated in Sec. VI. The results are discussed in Sec. VII. A summary and outlook are provided in Sec. VIII.

## II. REQUIREMENTS FOR A LOW-CODE ASSISTANCE SYSTEM

Complexity is a major obstacle to the understandability and maintainability of software [14]. To make software complexity explicit and thus controllable, an approach for quantifying complexity in VPL during programming shall be developed *(R1 complexity measurement)*.

One driver for high complexity in VPL projects is that repetitive code structures (*code clones*) are introduced via *copy & paste* to save time during programming, which may inflate a project. The assistance system shall automatically identify code clones in VPL and encapsulate them as a reusable unit *(R2 clone encapsulation)*.

The growing global competition requires a shorter time-to-market to stay competitive, increasing the pressure on programmers to develop more software in less time. Thus, the assistance system shall support programmers to develop software faster *(R3 time saving)*.

A popular means of assistance systems for textual languages to save time is the generation of recommendations for the next step intended by the programmer (e.g., auto-completion of variable names). Analogously, the proposed assistance system shall generate recommendations on which block could be needed next *(R4 recommendations)*.

The ever-increasing software complexity puts pressure on domain experts with little programming background, whose technical expertise, however, is a core requirement for correct software. Therefore, domain experts are the target group of the assistance system *(R5 support for domain experts)*.

To generate actual benefits with the assistance system during daily programming practice, it is required that the assistance system is intuitive to apply *(R6 intuitiveness)*.

Code analysis is a powerful lever for quality optimization of automation software, but usually requires time and additional steps in the workflow. Thus, the analysis results are often not further used for software optimization [15]. Therefore, it shall be possible to use the assistance system live during programming *(R7 online assistance)*.

## III. STATE OF THE ART

The state of the art in complexity measurement, code clone identification, and approaches for programming assistance systems are outlined in the following.

### A. Complexity Measurement of VPL

Static code analysis and software metrics are a valuable means of identifying optimization potentials without executing the code and measuring specific software characteristics, e.g., complexity [16] (cf. R1). However, static code analysis is not yet widespread for VPL, and existing approaches are often tailored to a specific language (e.g., commercial tools such as MathWork's *Model Metrics* [17] for Simulink). Plauska and Damaševičius [8] determine the complexity of VPLs using metrics and validate the approach on the VPLs *Lego NXT-G* and *Microsoft VPL*, but only individual blocks are examined and not complete programs [8]. Nickerson [18] measures the complexity of a VPL using, among others, a metric based on Halstead [19], i.e., different complexity measures based on the number of operands and operators in the code. Established metrics from computer science such as Halstead's metrics [19] or McCabe's cyclomatic complexity [20], which considers the number of decisions in a program, have also been successfully adapted for VPL in automation, e.g., to Ladder Diagram and Function Block Diagram defined by the IEC 61131-3 [14, 21]. Besides the structural program composition, also the layout quality strongly influences a VPL program's complexity, i.e., the visual arrangement of blocks and their connections. Taylor et al. [22] propose ten metrics to quantify the graphical design quality, e.g., edge crossing or symmetry.

In summary, there are multiple approaches to quantify the complexity of VPL but often tailored to specific VPL. Assistance for the programmer to reduce high complexity values is often not provided (cf. R2, R4, R5), and metrics alone are not sufficient to save time during programming (cf. R3). While most approaches can be integrated into the software development workflow, users are typically not supported in interpreting the results (R6), and, usually, additional steps are necessary to see the results, which is an obstacle for use in industrial practice (cf. R7).

### B. Clone Detection in VPL

One of the first algorithms for finding clones in graph-based modeling languages (cf. R2) is *CloneDetective* [23], which can be adapted to different textual programming languages and also to low-code platforms such as Simulink. However, this requires writing a suitable translator for each VPL. Approaches such as *ModelCD* [24] and *SIMONE* [25] find clones in Simulink models by converting the models to text and then to tokens as an abstract code representation to apply clone finding algorithms. For IEC 61131-3-compliant software, [26] apply a graph-mining approach to the software's call graphs to search for repetitive structures. Jnanamurthy et al. [27] conducted a semantic analysis of the dependencies of inputs and outputs of different blocks and [28] investigate clones by comparing metrics for different software variants in both graphical and textual languages.

Although several approaches exist to identify clones in VPL and graphical code representations, this is usually not possible online during programming (cf. R7). Most approaches place little emphasis on the intuitive interpretability of the analysis results making the application difficult for domain experts without deep software knowledge (cf. R4, R5, R6). Additionally, existing approaches often do not measure the success (e.g., reduction of complexity (cf. R1), time savings (cf. R3)) by encapsulating a clone.

### C. Data Mining and Assistance Systems for Programming

In industrial practice, assistance systems that provide recommendations for the elements to be used next (cf. R4) have been common for a long time for textual languages, e.g., Visual Studio's *ReSharper* [29]. Most of them are based on data mining approaches, e.g., to learn from existing projects. Data mining is often used as a synonym for *Knowledge Discovery in Databases (KDD)*, i.e., a five-step process including the *selection* of data, their *pre-processing*, the *transformation* into a suitable analysis format, the actual *data mining* for noticeable patterns, and the *interpretation* of the results. Using *Association Rule Mining (ARM)*, rules from graph-based data consisting of nodes and edges can be extracted to make predictions in a new data set based on the *confidence*, i.e., the relative frequency of how often a rule is true in the data set [30]. One ARM method is the *Frequent Subgraph Mining (FSM)* [31]. FSM can be *graph-transaction-based* (searching many graphs for repeating subgraphs) or *single-graph-based* (searching a single graph). One of the most frequently applied FSM algorithms is *gSpan* [32].

For textual languages, Proksch et al. [33] use data mining to derive recommendations for programmers by considering the context, the call frequency to existing methods, and rules derived from existing projects. Further approaches to generate recommendations to complete code are based, e.g., on natural language processing [34] or neural networks [35, 36].

For low-code and VPL, [37] develop a system to complete graphs by using grammars for graphs that specify what a correct graph may look like. *VisComplete* [38] is designed to complete VPLs based on existing projects that are searched for repetitive paths. The *SimVMA* system for Simulink [39] predicts complete systems based on partially implemented systems and generates next steps as recommendations. Deng et al. [40] provide a recommendation approach based on the analysis of subgraphs in existing projects. On this basis, a structural table is created that includes the subgraph leading to a selected node, the possibly following nodes, and the confidence for each combination. After the user selects a node, the similarity of the current subgraph is calculated for all subgraphs in the structure table. Potential recommendations are then sorted by similarity and confidence. Contrary to textual languages, commercial assistance systems for VPL to support the programmer by providing recommendations are still rare (R4). Low-code platforms such as Siemens Mendix [7] allow the simple programming of applications, but only little intuitively understandable assistance for domain experts during programming is provided (R5, R6, R7).

In summary, many concepts exist to support programmers during creating the software by recommendations. Some of these approaches use thereby also older projects to extract knowledge from. However, most of the approaches do not assist in reducing complexity (cf. R1, R2), and the potential time savings are not documented (R3).

To bridge this gap, this letter introduces an assistance system that provides recommendations, measures complexity, and encapsulates code clones during programming by combining static code analysis and data mining.

## IV. ASSISTANCE SYSTEM TO SUPPORT PROGRAMMERS TO DEVELOP LOW-CODE

To allow users to measure software complexity during coding (R1), find and replace clones within a project (R2), and obtain recommendations for blocks to be used next (R4), a concept for an assistance system is introduced comprising two phases (cf. Fig. 1): Before programming a new project (*Pre-Programming*), knowledge is extracted from existing projects (cf. Sec. IV.A), which is used for different kinds of assistance to support users online *during programming* (cf. Sec. IV.B).

### A. Pre-Programming – Analysis of Existing Projects

The steps of transforming VPL programs to enable data analysis, the selection of complexity metrics, and the creation of the database to derive recommendations are introduced.

#### 1) Selection and Transformation of Available Software Projects for Data Mining

The KDD process is followed to extract information from existing VPL projects. Accordingly, a *selection* of data is required first. The higher the number of available VPL programs, the more domain-specific information can be extracted, and the better is the quality of the recommendations afterwards. These projects are *pre-processed* by removing faulty or empty projects and *transformed* into a suitable representation to apply data mining methods. A graph-based representation is chosen, i.e., variables and blocks are represented as nodes and their interconnections as edges (cf. Fig. 2), motivated by two reasons: First, most VPLs already have a graph-like form, which makes the transformation easy, and a graph-based assistance system concept can thus be applied to almost all VPL. Second, graphs enable the application of a wide range of powerful analysis approaches since many formal methods from code analysis (e.g., for clone identification) and data mining (e.g., FSM) require a graph representation of the underlying data. When converting a VPL to a graph, the characteristics of the original programming language must be considered. For VPLs in which the data flows in a defined direction, e.g., the representation should be a directed graph to preserve the property of the flow direction. Parameter passing via edges can be considered in the form of additional nodes (one node per parameter passed) between the two original nodes. However, the more fine-grained the information of the original VPL structure is represented in the graphs, the more projects are required during the pre-processing to achieve reliable recommendations.

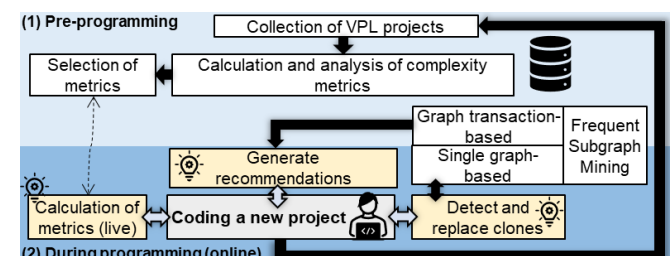

Fig. 1. Overview of the assistance system for programming low-code

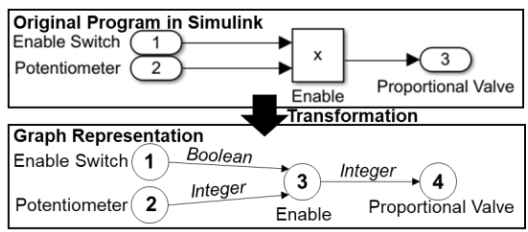

Fig. 2. Transformation of an exemplary Simulink code snippet to a graph representation consisting of nodes and (directed) edges.

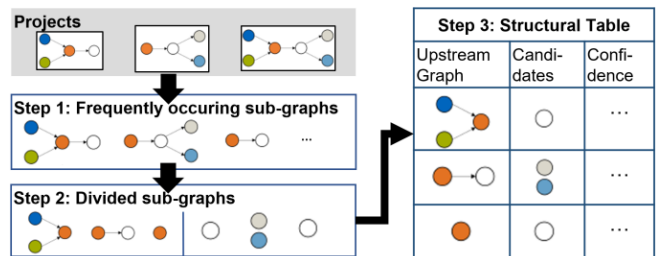

Fig. 3. Steps to generate the structural table as basis for recommendations.

*2) Selection of Complexity Metrics*

Software metrics are an effective means to assess complexity and, thus, ideal for addressing R1. First, a large set of metrics that can be principally applied to VPL is initially computed (cf. Sec. III.A). The results are gathered in a metrics table for each project. A two-step analysis is performed: *First*, the variance of each metric is calculated across all projects to determine how much a particular characteristic of the programs varies. Metrics with little or no variance are not suitable for comparing different projects and are therefore discarded. *Second*, an analysis according to [41] is performed. Bravais-Pearson's correlation coefficients are calculated to identify the extent to which two values are linearly dependent, i.e., redundant for the complexity measurement.

The decision on which of the highly correlated metrics to use should be made by a domain expert from the target group to choose the most intuitive metrics, which cannot be achieved with a purely mathematic selection. The metric selection does require effort, but only a single time during pre-processing, and the selection can then be used repeatedly for analysis or readjusted as needed. In the long run, also a pre-defined set of metrics could be recommended by the platform supplier, e.g., based on experience, which metrics are often used by customers. The result of the selection process is a list of non-redundant metrics that measure only characteristics that vary across the analyzed projects.

*3) Database for Recommendations*

Data-driven approaches have the advantage that no precise knowledge about the VPL (e.g., its grammar) is needed since the recommendations are based only on the available data. The approach used for the concept is an extension of [40]. Using a graph-transaction-based FSM algorithm based on [32], possible repetitive structures are first extracted from the graphs (cf. Fig. 3). The most important parameter for the search for subgraphs is the support, i.e., the minimum number of projects in which a subgraph must be contained. The support should allow finding as many structures as possible to generate the highest possible number of rules and, at the same time, to find only those structures that are frequently used to avoid overfitting (step 1). A recurrent structure should contain at least two nodes and one edge to allow a later separation into upstream subgraph (i.e., what has been programmed) and candidates (i.e., blocks that may come next) (step 2). In step 3, the results are stored in a structural table, and the confidence is calculated for each row, indicating how likely these candidates are to follow the respective upstream graph.

*B. During Programming – Online Assistance*

The following sub-sections describe how the assistance system concept supports the user online during programming based on the data set established in phase 1.

*1) Online Calculation of Complexity Metrics*

The calculation of complexity metrics online during programming is achieved by having a graph representation for the current program available at all times. The metrics are recalculated with each change, thus providing the user with direct feedback on how a change affects the complexity. The selected metrics should be directly visible in a user interface and not hidden in submenus. To avoid overwhelming the user, only a small set of metrics is displayed directly. Therefore, the metrics selected in Sec. IV.B.1 may be further reduced.

*2) Online Recommendations*

The process for generating recommendations is based on the structural table (cf. Fig. 3) and consists of four steps (cf. Fig. 4). After the user selects an element (node) for which recommendations shall be generated, the upstream graph leading to the selected node is calculated, which serves as the reference graph to identify candidates for blocks to use next (Step 1). In Step 2, the similarity between the reference graph and all upstream graphs is stored in the structural table. The similarity values are based on the *Graph Editing Distance (GED)*, i.e., the minimum number of steps (e.g., adding or deleting nodes) to transfer one graph to another [42]. In Step 3, the confidence of all rules, including the same candidates, is summed up, and the merged rule is assigned the minimum GED value for the corresponding upstream subgraphs. All summarized rules are sorted in ascending order according to the minimum GED. Next, a number of recommendations is displayed to the user to be used for programming (Step 4).

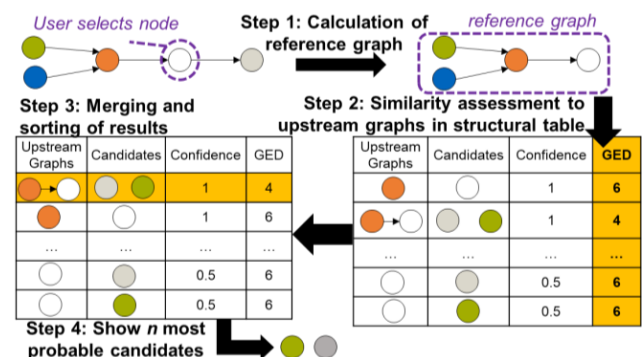

Fig. 4. Steps to derive recommendations for blocks to use next.

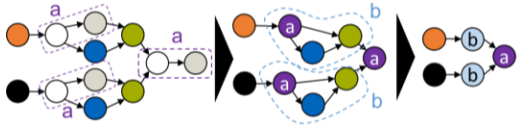

Fig. 5. Process for automatic encapsulation of repeating subgraphs.

*3) Online Replacement of Clones in the Project*

A single-graph-based FSM approach is used to extract repetitive structures from the program. Subgraphs that are more frequently repeated are encapsulated first to allow for the quickest and easiest encapsulation of projects. If two subgraphs occur equally often in a reference graph, the larger one is prioritized to minimize the number of steps needed to simplify the project (cf. Fig. 5).

## V. IMPLEMENTATION OF THE ASSISTANCE SYSTEM

In the following, the assistance system's implementation for a concrete VPL to evaluate its benefits for supporting users during programming, is introduced.

### A. Use Case and Selection of a VPL

Hydraulic components are used for numerous applications in automation and manufacturing, such as mobile machines (e.g., in agriculture) or stationary equipment (e.g., machine tools and presses). The digitalization of intelligent hydraulic components is a key enabler in providing services as part of Industry 4.0 [13], leading to an increasing software complexity. Since the rising amount of software to control hydraulic actuators pushes domain experts to carry out more programming work, e.g., to integrate hydraulic components into existing machines, domain-specific low-code platforms become increasingly important. Therefore, the assistance system is implemented for the industrial low-code platform eDesign [12] developed by HAWE Hydraulik SE, a hydraulic components and platform supplier. eDesign facilitates the programming of hydraulic components with pre-defined function blocks connectable via ports (cf. Fig. 6). In the case of web-based low-code platforms being the target of this approach, it is assumed that suppliers often have access to a large number of customer projects from a certain domain, which can be used for pre-processing. For this letter, 1269 anonymized eDesign projects are provided by HAWE.

### B. Complexity Metrics for the Selected Language

For complexity measurement, the most common metrics that have proven to be suitable for VPL programs [14] are used for the prototype as default, i.e., McCabe's Cyclomatic Complexity, Halstead's Length, Vocabulary, and Difficulty. eDesign allows the user to arrange the blocks freely and, thus, influence the program's layout quality. Therefore, layout metrics [22] are implemented and weighted (cf. Tab. 1).

While the structural complexity of two programs may be the same, and thus metrics such as cyclomatic complexity do not change between the two programs, the layout quality and, thus, the program's understandability may vary, which can be made explicit using the selected layout metrics (cf. Fig. 6).

TABLE I. WEIGHTED LAYOUT METRICS ACCORDING TO [22]

| Metric | Weight |
|---|---|
| Angular Resolution | $1 \cdot 10^{-2}$ |
| Aspect Ratio | $1 \cdot 10^{-7}$ |
| Edge Overlaps | 1 |
| Nearest Neighbour Variance | $1 \cdot 10^{-6}$ |
| Uniform Edges | $1 \cdot 10^{-3}$ |
| Concentration | 1 |
| Homogeneity | 1 |

### C. Description of the Assistance System Prototype

The implementation of the assistance system is developed in C# and allows loading eDesign projects and calculating the selected metrics automatically during the import. The user interface (cf. Fig. 7) supports the user during programming in different sub-areas. Once a new project has been loaded or created (area 1), area 2 enables programming in a VPL that visually corresponds to eDesign. Area 3 contains blocks defined in eDesign and can be inserted into area 2 via *drag & drop*. Area 4 provides assistance features to reduce the program's complexity, i.e., automated encapsulation of recurring sub-graphs as blocks that can be reused (then stored in area 3) and layout optimization. The impact of the complexity reduction is quantified by the selected metrics displayed in area 5. The values are automatically recalculated as soon as the user changes the program. Additional metrics such as the individual values of the overall layout quality can be displayed in area 6 for more details. In area 7, the candidates of the recommendation system are displayed. These are calculated when the user selects a block in area 2.

## VI. EVALUATION OF THE ASSISTANCE SYSTEM

The evaluation of the assistance system is carried out in two stages. First, a user study is conducted with a suitable target group (cf. Sec. VI.A). Second, an industrial expert workshop with HAWE employees responsible for eDesign is held to assess the potential of the assistance system for industrial application by their users (cf. Sec. VI.B).

### A. User Study with Domain Experts

The user study's design, implementation, and results are introduced in the following.

*1) Design of the user study*

Companies using domain-specific low-code platforms such as eDesign usually have only a few experts working with the platform impeding representative industrial user studies. However, previous work [43] shows that student evaluations may yield relevant results when substantiated with further evaluation elements such as industrial interviews (cf. Sec. VI.A.2). The assistance system is, therefore, evaluated in a first step with students whose background is as close as possible to industrial domain experts, i.e., in-depth knowledge in automation and hydraulics and little experience in software engineering (R5). Ten students from mechanical engineering, mechatronics, and robotics were selected.

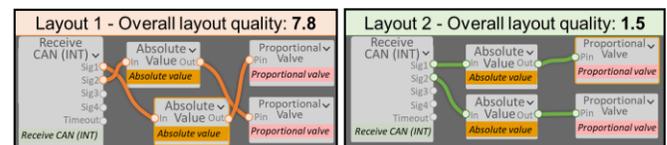

Fig. 6. Comparison of the layout quality of two structurally identical programs in eDesign (Cyclomatic Complexity [17] = 1 in both cases).

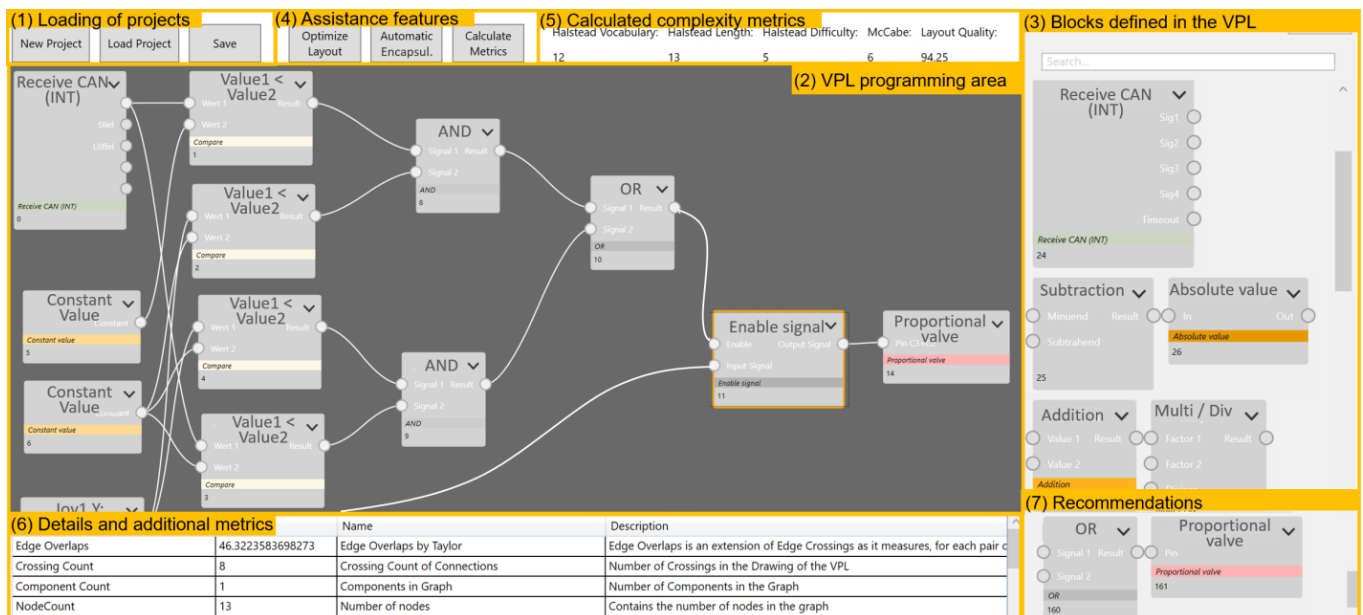

Fig. 7. Overview of the user interface of the prototypical implementation of the assistance system for the low-code platform eDesign.

Two slightly modified versions of the user interface are used: Group 1 uses the complete assistance system (cf. Fig. 7). For Group 2, the assistance features (areas 4, 5, and 7) are disabled, thus simulating the original programming environment of eDesign. After a brief introduction, the participants fill out a questionnaire about their area of expertise and programming skills. Subsequently, they familiarize themselves with the user interface used in their group. Next, the participants solve programming tasks, which stem from an example project to control the motion path of an excavator arm, i.e., a typical task for component integrators in industrial practice (cf. Tab. 2). For *Tasks 1-3*, the time needed is measured. In *Task 4*, the participants rate the complexity of five given projects. To evaluate whether encapsulating blocks reduces the complexity, the results of Task 3 of Group 1 (with encapsulation) and Group 2 (without encapsulation) are also included. While Group 1 is finished after Task 4, Group 2 switches to the user interface with assistance features to be tested in a *Task 5,* in which the results of Task 2 are encapsulated and reused in a given project. Finally, all participants rate the assistance in a second questionnaire.

*2) Results of the user study*

While no time savings using the assistance system could be measured for Task 1 (explainable by the simplicity of the task and an initial effort for familiarization), significant time was saved for Task 2 (22% less time) and Task 3 (54% less time) using the recommendations and the encapsulation of blocks (R3). Most participants in Group 1 used the recommendations for programming, which proves that they are helpful (R4). The complexity assessment of the projects (Task 4) strongly coincides with the metrics results proving their capability to quantify complexity (R1) and shows that the complexity of programs is reduced using encapsulated blocks (R2). Since the participants partly used the assistance features subconsciously and automatically, good intuitiveness (R6) and the applicability during programming (R7) are confirmed.

These observations are confirmed by the questionnaire (cf. Fig. 8). The encapsulation of blocks is consistently rated positively. Additionally, it is confirmed that encapsulation can reduce complexity (R2). The participants rate the assistance features' intuitiveness as very good (R6), so it can be concluded that the concept is helpful for the target group (R5). Nevertheless, some participants are unsure how often they would use the recommendations. While 60% of the participants state that displaying complexity metrics at least partially helps write less complex programs, 40% rather disagree with this statement, i.e., not all participants consider displaying the metrics helpful but tend to agree with the statement (R1). Overall, all participants agree that the assistance system can be used during programming (R7).

TABLE II. DESCRIPTION OF TASKS DURING EVALUATION (G = GROUP)

| Task | Description | Type of task | Performed by | | Assistance enabled | |
|---|---|---|---|---|---|---|
| | | | G1 | G2 | G1 | G2 |
| 1 | Add and connect two blocks to familiarize with the user interface | Programming with time tracked | x | x | x | - |
| 2 | Create a project with multiple operators to be reused in later tasks | Programming with time tracked | x | x | x | - |
| 3 | Insert results of Task 2 a second time into the project | Programming with time tracked | x | x | x | - |
| 4 | Rate complexity of five given projects from 1-10. | Questionnaire | x | x | - | - |
| 5 | Encapsulate and reuse the result of Task 2 in a given project | Programming | - | x | n.a. | x |

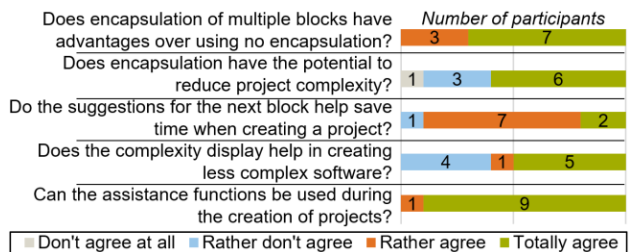

Fig. 8. Overview of questionnaire results obtained in the user study.

### B. Industrial Expert Workshop

To substantiate the user study and assess the assistance system's potential for industrial application, a workshop with three platform developers from HAWE responsible for eDesign is conducted.

*1) Design of the Workshop*

First, the assistance system's functionalities are introduced. Next, the assistance features are demonstrated using example projects, i.e., the complexity calculation of the loaded project, the encapsulation and reuse of blocks, and the generation of recommendations. The strengths and weaknesses, and the applicability of the assistance system, are reviewed with the experts using a questionnaire and a group discussion.

*2) Workshop Results*

The expert workshop confirmed that the scaling-up problem is at least partially a challenge for the target group of eDesign and, thus, confirms the need for assistance to reduce complexity. The selected metrics quantify complexity in an understandable way (R1). According to the experts, block encapsulation has a great potential to reduce the project complexity (R2) since the encapsulation leads to a better overview and thus to less complexity (mentioned by all), and may help to become aware of recurring structures, thus encouraging more structured thinking. The experts expect that the recommendations can at least partly save time (R3) and are helpful for inexperienced users to provide ideas for more complex issues (R4). The automatic encapsulation of clones, e.g., can accelerate the refactoring of legacy projects (R3). The approach based on data from existing projects is rated as helpful to extract knowledge, which would be difficult to do manually. In summary, the assistance is perceived by all experts as helpful for the industrial target group of eDesign (R5), and the questionnaire confirmed that the features are expected to be applicable during programming (R7).

## VII. DISCUSSION OF RESULTS AND THREATS TO VALIDITY

Group 2 used an adapted user interface with deactivated assistance features instead of the original eDesign environment for better comparability of both groups during the study. This allows valid statements about the assistance with the implemented user interface, and the expert workshop with HAWE confirmed the transferability to the original programming environment, which is why the comparison results are assessed as valid.

The assessment of time savings using the assistance system is limited since copying large structures instead of reusing encapsulated blocks is a common way to save time on short notice yet with the risk of time-consuming maintenance in the future. Assessing these short- and long-term effects on time-efficient programming needs to be targeted in future work.

One exemplary low-code platform, i.e., eDesign, was focused in the evaluation. It is assumed that the conceptual features, e.g., measuring complexity or encapsulation, can be transferred to other low-code platforms since they can be applied to all graph-like VPL. However, usability is dependent on the particular platform. To avoid this effect, the questions focused on the features per se and not their implementation, but a residual influence of the implementation on the respondents' perception remains. In future work, also the scalability needs to be tested for larger data sets and different platforms, as the complexity of data-based approaches increases over-proportionally with the amount of data.

The selection of participants may influence the validity of the results. In this case, students were selected having similar knowledge as the industrial target group. The transferability to the target group is assessed as valid since the interviewed platform developers confirmed the applicability and usefulness for their customers. However, this expectation needs to be validated with industrial domain experts, e.g., with evaluations in selected focus groups from the industrial user community together with the platform supplier.

In summary, the overall validity of the results is considered high and promising (cf. Tab. 3). Future work needs to explore implementing the concept for other low-code platforms with additional user groups

## VIII. CONCLUSION AND OUTLOOK

An assistance system to support domain experts in developing low-code by measuring complexity and encapsulating clones to reduce the scaling-up problem in VPL is introduced. Using data mining, recommendations for blocks to be used are automatically identified to support programmers in efficiently writing software to cope with increasing time pressure. The benefits of the system are validated in a user study and an industrial expert workshop with a prototypical implementation for the low-code platform eDesign based on an analysis of more than 1200 user projects, indicating the

TABLE III. REQUIREMENT FULFILLMENT AND VALIDITY (🟢 = FULLY FULFILLED / HIGH VALDIDITY; 🟡 = PARTLY FULFILLED / MEDIUM VALIDITY)

|  | Insights | Result | Validity |
|---|---|---|---|
| R1 | Selected metrics are capable of quantifying complexity; values are intuitively interpretable | 🟢 | 🟢 |
| R2 | Automatic encapsulation of clones significantly reduces complexity | 🟢 | 🟢 |
| R3 | Assistance system significantly saves time by providing recommendations and encapsulation (more than 50% measured) | 🟢 | 🟡 |
| R4 | Generated recommendations based on data mining are accepted and used during programming | 🟢 | 🟢 |
| R5 | Helpfulness of assistance system for domain experts with little programming background confirmed in user study | 🟢 | 🟢 |
| R6 | Subconscious use of assistance features in user study confirms intuitiveness | 🟢 | 🟡 |
| R7 | Applicability during programming confirmed in user study and expert workshop | 🟢 | 🟢 |

This article has been accepted for publication in IEEE Robotics and Automation Letters. This is the author's version which has not been fully edited and content may change prior to final publication. Citation information: DOI 10.1109/LRA.2022.3193728

IEEE ROBOTICS AND AUTOMATION LETTERS. PREPRINT VERSION. ACCEPTED JULY, 2022potential of the assistance system to be implemented by platform suppliers in the future to support industrial domain experts in performing programming tasks, such as integrating hydraulic components into mobile machines.

Research in the field of Industry 5.0 clearly shows the importance of supporting humans through automation in the context of ever shorter innovation cycles and the increasing complexity of mechatronic systems, especially in software. As an increasing amount of complex functionalities of mechatronic systems, which require in-depth knowledge of the technical system, is implemented in the software, the role of domain experts in software development becomes steadily more important. Thus, the relevance of low-code platforms will continue to grow. Implementations of the assistance system concept for additional low-code platforms for user groups with different qualification levels are planned in the future, as well as further industrial evaluations with focus groups of the user community in cooperation with low-code platform suppliers to enhance and generalize the approach for different domains. To not only support users but also motivate them, aspects such as gamification elements during the reduction of software complexity shall be considered.

## REFERENCES

[1] S. Nahavandi, "Industry 5.0—A Human-Centric Solution," *Sustainability*, vol. 11, no. 16, p. 4371, 2019.
[2] X. Xu, Y. Lu, B. Vogel-Heuser, and L. Wang, "Industry 4.0 and Industry 5.0—Inception, conception and perception," *Journal of Manufacturing Systems*, vol. 61, pp. 530–535, 2021.
[3] MathWorks, *Simulink*. [Online]. Available: https://de.mathworks.com/products/simulink.html (accessed: Feb. 12 2022).
[4] *IEC 61131 Programmable controllers - Part 3: Programming languages*, International Electrotechnical Commission, 2013.
[5] B. Jost, M. Ketterl, R. Budde, and T. Leimbach, "Graphical Programming Environments for Educational Robots: Open Roberta - Yet Another One?," in *IEEE ISM*, 2014, pp. 381–386.
[6] K. N. Whitley and A. F. Blackwell, "Visual programming," in *7th Workshop on Empirical Studies of Programmers*, 1997, pp. 180–208.
[7] Siemens, *Mendix*. [Online]. Available: https://www.plm.automation.siemens.com/global/de/products/mendix/ (accessed: Feb. 10 2022).
[8] I. Plauska and R. Damaševičius, "Usability analysis of visual programming languages using computational metrics," *IADIS Interfaces and Human Computer Interaction (IHCI)*, pp. 63–70, 2013.
[9] C. Richardson and J. R. Rymer, "Vendor landscape: The fractured, fertile terrain of low-code application platforms," *FORRESTER, Janeiro*, 2016.
[10] R. Sanchis, Ó. García-Perales, F. Fraile, and R. Poler, "Low-Code as Enabler of Digital Transformation in Manufacturing Industry," *Applied Sciences*, vol. 10, no. 1, p. 12, 2020.
[11] I. Burnett, M. J. Baker, C. Bohus, P. Carlson, S. Yang, and P. van Zee, "Scaling up visual programming languages," *Computer*, vol. 28, no. 3, pp. 45–54, 1995.
[12] HAWE Hydraulik SE, *eDesign - The Graphical Programming Interface for Hydraulic Controls*. [Online]. Available: https://www.hawe.com/topics/hawe-edesign/ (accessed: Feb. 28 2022).
[13] VDMA, "Fluid Power 4.0 – digitize, connect, communicate," 2019. Accessed: May 26 2022. [Online]. Available: https://vdma.org/viewer/-/v2article/render/1187414
[14] J. Fischer, B. Vogel-Heuser, H. Schneider, N. Langer, M. Felger, and M. Bengel, "Measuring the Overall Complexity of Graphical and Textual IEC 61131-3 Control Software," *IEEE RA-L*, pp. 5784–5791, 2021.
[15] M. Beller, R. Bholanath, S. McIntosh, and A. Zaidman, "Analyzing the State of Static Analysis: A Large-Scale Evaluation in Open Source Software," in *IEEE 23rd SANER*, 2016, pp. 470–481.
[16] H. Prähofer, F. Angerer, R. Ramler, H. Lacheiner, and F. Grillenberger, "Opportunities and challenges of static code analysis of IEC 61131-3 programs," in *Proc. of IEEE 17th ETFA*, 2012, pp. 1–8.
[17] MathWorks, *Model Metrics*. [Online]. Available: https://de.mathworks.com/help/slcheck/model-metrics.html (accessed: Feb. 10 2022).
[18] J. V. Nickerson, "Visual programming: limits of graphic representation," in *Proc. of IEEE Symposium on Visual Languages*, 1994, pp. 178–179.
[19] M. H. Halstead, *Elements of software science*. New York and Oxford: Elsevier, 1977.
[20] T. J. McCabe, "A Complexity Measure," *IEEE Transactions on Software Engineering*, SE-2, no. 4, pp. 308–320, 1976.
[21] L. Capitán and B. Vogel-Heuser, "Metrics for software quality in automated production systems as an indicator for technical debt," in *13th IEEE CASE*, 2017, pp. 709–716.
[22] M. Taylor and P. Rodgers, "Applying Graphical Design Techniques to Graph Visualisation," in *9th Int. Conf. on Information Visualisation*, 2005, pp. 651–656.
[23] E. Juergens, F. Deissenboeck, B. Hummel, and S. Wagner, "Do code clones matter?," in *IEEE 31st ICSE*, 2009, pp. 485–495.
[24] N. H. Pham, H. A. Nguyen, T. T. Nguyen, J. M. Al-Kofahi, and T. N. Nguyen, "Complete and accurate clone detection in graph-based models," in *IEEE 31st ICSE*, 2009, pp. 276–286.
[25] J. R. Cordy, "Submodel pattern extraction for simulink models," in *Proc. of 17th SPLC*, 2013, p. 7.
[26] M. Fahimipirehgalin, J. Fischer, S. Bougouffa, and B. Vogel-Heuser, "Similarity Analysis of Control Software Using Graph Mining," in *IEEE 17th INDIN*, 2019, pp. 508–515.
[27] H. K. Jnanamurthy, R. Jetley, F. Henskens, D. Paul, M. Wallis, and S. D. Sudarsan, "Analysis of Industrial Control System Software to Detect Semantic Clones," in *IEEE ICIT*, 2019, pp. 773–779.
[28] K. Rosiak, A. Schlie, L. Linsbauer, B. Vogel-Heuser, and I. Schaefer, "Custom-tailored clone detection for IEC 61131-3 programming languages," *Journal of Systems and Software*, vol. 182, pp. 1–18, 2021.
[29] M. Koch, *Inspections and Quick-Fixes in ReSharper*. [Online]. Available: https://www.jetbrains.com/dotnet/guide/tutorials/resharper-essentials/inspections-quick-fixes/ (accessed: Feb. 4 2022).
[30] J. Cleve and U. Lämmel, *Data Mining*: De Gruyter, 2014.
[31] S. C. Dimri, P. Malik, and M. Ram, *Algorithms*: De Gruyter, 2021.
[32] X. Yan and J. Han, "gSpan: graph-based substructure pattern mining," in *IEEE Int. Conf. on Data Mining*, 2002, pp. 721–724.
[33] S. Proksch, J. Lerch, and M. Mezini, "Intelligent Code Completion with Bayesian Networks," *ACM TOSEM*, vol. 25, no. 1, 2015.
[34] V. Raychev, M. Vechev, and E. Yahav, "Code completion with statistical language models," in *Proc. of 35th ACM SIGPLAN PLDI*, 2014, pp. 419–428.
[35] A. Svyatkovskiy, Y. Zhao, S. Fu, and N. Sundaresan, "Pythia: AI-assisted Code Completion System," in *25th ACM SIGKDD KDD*, 2019, pp. 2727–2735.
[36] M. S. Kalyon and Y. S. Akgul, "A Two Phase Smart Code Editor," in *3rd HORA*, 2021, pp. 1–4.
[37] S. Mazanek, S. Maier, and M. Minas, "Auto-completion for diagram editors based on graph grammars," in *IEEE VL/HCC*, 2008, pp. 242–245.
[38] D. Koop, C. E. Scheidegger, S. P. Callahan, J. Freire, and C. T. Silva, "VisComplete: automating suggestions for visualization pipelines," *IEEE TVCG*, vol. 14, no. 6, pp. 1691–1698, 2008.
[39] M. Stephan, "Towards a Cognizant Virtual Software Modeling Assistant using Model Clones," in *IEEE/ACM 41st ICSE-NIER*, 2019, pp. 21–24.
[40] S. Deng et al., "A Recommendation System to Facilitate Business Process Modeling," *IEEE transactions on cybernetics*, vol. 47, no. 6, pp. 1380–1394, 2017.
[41] I. Herraiz and A. E. Hassan, "Beyond lines of code: Do we need more complexity metrics?," *Making software: what really works, and why we believe it*, pp. 125–141, 2010.
[42] M. E. J. Newman, *Networks: An introduction / M.E.J. Newman*. Oxford: Oxford University Press, 2010.
[43] M. Obermeier, S. Braun, and B. Vogel-Heuser, "A Model-Driven Approach on Object-Oriented PLC Programming for Manufacturing Systems with Regard to Usability," *IEEE TII*, vol. 11, no. 3, pp. 790–800, 2015.© 2022 IEEE. Personal use is permitted, but republication/redistribution requires IEEE permission. See https://www.ieee.org/publications/rights/index.html for more information.
Authorized licensed use limited to: Technische Universitaet Muenchen. Downloaded on July 26,2022 at 11:01:35 UTC from IEEE Xplore. Restrictions apply.